\documentclass[final,5p,times]{elsarticle}
\usepackage{graphicx}
\usepackage{amsmath}
\begin{document}
\newcommand{\op}{\boldsymbol}
\journal{Annals of Physics}
\begin{frontmatter}

\title{Wave-Particle Duality in N-Path Interference}
\author[label1]{Tabish Qureshi}
\ead{tabish@ctp-jamia.res.in}
\address[label1]{Centre for Theoretical Physics, Jamia Millia Islamia, New Delhi-110025, India.}
\author[label2,label1]{Mohd Asad Siddiqui}
\ead{asad@ctp-jamia.res.in}
\address[label2]{Department of Physics,
	National Institute of Technology,
	Patna-800005, India.}

\begin{abstract}

Bohr's principle of complementarity,
in the context of a two-slit interference experiment, is understood as
the quantitative
measures of wave and particle natures following a duality relation
${\mathcal D}^2+{\mathcal V}^2 \le 1$. Here ${\mathcal D}$ is a measure of
distinguishability of the two paths, and ${\mathcal V}$ is the visibility of
interference.
It is shown that such a relation can be formulated for $N-$slit or
$N-$path interference too, with the proviso that the wave nature is 
characterized by a measure of {\em coherence} (${\mathcal C}$). This new relation,
${\mathcal D}^2+{\mathcal C}^2 \le 1$ is shown to be tight, and reduces to the
known duality relation for the case $N=2$. A recently introduced
similar relation (Bagan et al., 2016) is shown to be inadequate for the purpose.


\end{abstract}


\end{frontmatter}





\section{Introduction}

Niels Bohr had argued that the wave and particle natures of quantum objects
are complementary \cite{bohr}. An experiment that clearly illustrates 
one nature, will necessarily obscure the other. Quantum objects, which
show the properties of both a particle and a wave are often called
quantons \cite{bunge,levy}.
In a two-slit interference experiment, the particle nature is characterized by
the ability to tell, which of the two slits the quanton went through.
The wave nature, on the other hand, has traditionally been characterized by
the interference patterns built up by the successive registering of individual
quantons on a screen. The visibility of the interference pattern is a 
good measure of the wave nature. 

An inequality was derived by Englert \cite{englert}, which can be understood
as a quantitative statement of Bohr's complementarity principle
\begin{equation}
{\mathcal D}^2 + {\mathcal V}^2 \le 1,
\label{englert}
\end{equation}
where  ${\mathcal D}$ is a {\em path distinguishability} and ${\mathcal V}$ is
the visibility of the interference pattern. One should note that ${\mathcal D}$ 
here does not itself have a physical meaning, although it is a good measure of
the ability to distinguish between the two paths of a quanton. This work built up
on various earlier attempts \cite{wootters,greenberger,jaeger}.
This duality is now well established, and has also been connected
to entropic uncertainty relation \cite{coles} and to a dichotomy between
symmetry and asymmetry \cite{vaccaro}, among others.

A natural question then arises, that since wave-particle duality should also
hold for multi-slit interference experiments, can a duality relation be
formulated for such experiments? Several attempts have been made in this
direction, particularly for three-slit interference experiments
\cite{durr,bimonte,bimonte1,englertmb,zawisky,hess}. Earlier we derived a duality
relation for three-slit interference \cite{3slit}, by introducing a new
distinguishability ${\mathcal D}_Q$ based on unambiguous quantum state discrimination
(UQSD) \cite{uqsd, dieks, peres, jaeger2, bergou}.

For N-path interference a duality relation ${\mathcal D}_Q + {\mathcal C} \le 1$
was derived by Bera et al. \cite{nslit}, where ${\mathcal D}_Q$ is a new
distinguishability based on UQSD, and ${\mathcal C}$ is a {\em coherence} based on 
a recently introduced measure of quantum coherence \cite{coherence}. This relation
has been shown to be tight, and also reduces to Eqn (\ref{englert}) for
$N=2$. In spite of this, there seems to be an interest in deriving a
relation for N-path interference, which has a form similar to (\ref{englert}) 
\cite{bagan,coles1}. 

In this paper we formulate a duality relation, 
\begin{equation}
{\mathcal D}^2 + {\mathcal C}^2 \le 1,
\label{nduality}
\end{equation}
which is in the form of (\ref{englert}), and holds for N-path interference.
For the case $N=2$, it reduces to (\ref{englert}), and the distinguishability
${\mathcal D}$ and coherence ${\mathcal C}$ reduce to Englert's 
distinguishability and visibility \cite{englert}, respectively.

\section{N-path interference}

We start by considering a N-path interference experiment. There are $N$
paths available for the quanton to pass through, before it encounters a
screen or detector to give rise to interference. We consider a general
scenario where the probabilities to pass through different paths (or slits)
may be unequal. The state of the quanton after passing through the 
$N$ paths may be written as 
$|\Psi\rangle = c_1|\psi_1\rangle+c_2|\psi_2\rangle+c_3|\psi_3\rangle
+ \dots + c_N|\psi_N\rangle$,
where $|\psi_i\rangle$ is the possible state of the quanton if it passes
through the i'th path (or slit), and $c_i$ is the amplitude for taking that
path. If $\{|\psi_i\rangle\}$ are orthonormal, $c_i$ should satisfy
$\sum_i |c_i|^2 = 1$.

Consider now a path-detector which is capable of recording which path the
quanton followed. This path detector is also a quantum object. According
to von Neumann's criteria of a quantum measurement \cite{neumann}, the
states of the quanton, corresponding to it passing through different paths,
should get correlated to distinct states of the path-detector.
We assume that the initial state of the quanton, as it passes through the
$N$ paths, is $|\Psi\rangle = \sum_{i=1}^N c_i|\psi_i\rangle$, and the 
path-detector is in a definite state $|d_0\rangle$.
A unitary operation, which corresponds to a von Neumann measurement interaction,
takes the combined state, of the quanton and the
path-detector, to an entangled state, which can be written as
\begin{equation}
|\Psi\rangle = c_1|\psi_1\rangle|d_1\rangle+c_2|\psi_2\rangle|d_2\rangle
+ \dots + c_N|\psi_N\rangle|d_N\rangle,
\label{corrstate}
\end{equation}
where $|d_i\rangle$ is the state of the path-detector corresponding to the
quanton following the i'th path. Without loss of generality, $|d_i\rangle$
can be assumed to be normalized, but not necessarily orthogonal.
Given the entangled state (\ref{corrstate}), the d-system states carry 
the path information about the quanton.
If the states in the set $\{|d_i\rangle\}$
are all orthogonal to each other, one could measure an operator of the
path-detector, which has different eigenvalues corresponding to different
$|d_i\rangle$'s. Getting a result, say, $|d_k\rangle$ would imply that the
quanton went through the k'th path.

\section{Path distinguishability}

On the other hand, if $\{|d_i\rangle\}$ are not all orthogonal to each other,
one cannot distinguish between various $|d_i\rangle$'s, and thus between
various quanton paths, with certainty. In such a situation, one needs to
define a {\em distinguishability} of the paths. If the quanton and 
path-detector state is pure, as given by (\ref{corrstate}), in Englert's
formulation, path distinguishability, for the case $N=2$, and
$|c_1|=|c_2|=1/\sqrt{2}$, comes out to be
\begin{equation}
{\mathcal D} = \sqrt{1 - |\langle d_1|d_2\rangle|^2}.
\label{Denglert}
\end{equation}
This ${\mathcal D}$, whose values are restricted to $0 \le {\mathcal D} \le 1$,
is a good measure of how well the paths can be distinguished.
Although it does not have a physical meaning in itself, ${\mathcal D}$ was
motivated through
minimum error discrimination of states.

For the case of $N$ paths, we introduce path distinguishability as
\begin{equation}
{\mathcal D} \equiv \sqrt{1 - \left({1\over N-1}\sum_{i\neq j} |c_ic_j| |\langle d_i|d_j\rangle|\right)^2}.
\label{Dtqas}
\end{equation}
Our proposed $\mathcal{D}$ satisfies the first four  basic criteria suggested by D\"urr for being reliable path quantifier\cite{durr}:
	
		 (1) $\mathcal{D}$ is a continuous function of the probabilities, $p_i=|c_i|^2$,
		 (2) $\mathcal{D}$ reaches its global maximum, when we have the perfect knowledge of the
		  path acquired by the quanton, (i.e $p_i=1$ for one beam, rest other $p_j=0$, and $\langle d_i|d_j\rangle's=0$),
	(3) $\mathcal{D}$   reaches its global minimum, when there is equal possibility for quanton to acquire any slit
	 ($p_i=1/n\ \forall i$), and all detectors are parallel ($\langle d_i|d_j\rangle's=1$), 
	(4) Any attempt towards the equalization of the probabilities
		 ($p_1, p_2, \dots p_n$) or parallelization of detector states ($|d_i\rangle, |d_i\rangle, \dots |d_n\rangle$),
		 will  decrease the measure of $\mathcal{D}$.
		 
		 For the case, $N=2$ and $|c_1|=|c_2|=1/\sqrt{2}$, it reduces to Englert's
distinguishability (\ref{Denglert}). 
 It can also be shown that ${\mathcal D}$,
given by (\ref{Dtqas}) is restricted to $0 \le {\mathcal D} \le 1$. One can see
that if $|d_i\rangle$'s are all mutually orthogonal, ${\mathcal D}$ is equal to 1,
which corresponds to full path-information.
We had earlier proposed a distinguishability for N-path interference, given by
\cite{3slit}
\begin{equation}
{\mathcal D}_Q \equiv 1 - {1\over N-1}\sum_{i\neq j} |c_ic_j| |\langle d_i|d_j\rangle|.
\label{Dbera}
\end{equation}
This distinguishability, which has been used by Bera et al. too, has
a clear physical meaning, which is explained in
the following. Suppose one is given an 
arbitrary state of the path-detector, and the question asked is which one
is it out of the set $\{|d_i\rangle\}$. One can answer this using UQSD. 
UQSD is a method in which the measurement results (on the given $d$-state,
in our case) can be divided into two categories. In one category, one can tell 
{\em without any error} which amongst the set $\{|d_i\rangle\}$ is the given
$d$-state. In the other category, one cannot tell which state it is at all.
In other words, if the distinguishing process succeeds, it does so without
any error, otherwise it fails. Maximizing the probability of success, and
thus minimizing the probability of failure, is what is 
important here. The distinguishability ${\mathcal D}_Q$
is an upper bound on the probability of successful discrimination. In other words, the states in the
set $\{|d_i\rangle\}$ cannot be distinguished {\em without any error} with
a probability larger than ${\mathcal D}_Q$ \cite{zhang,qiu}. This is the
sense in which ${\mathcal D}_Q$ has a physical meaning.
It is easy to verify that the distinguishability introduced in (\ref{Dtqas})
is related to ${\mathcal D}_Q$ by the relation 
${\mathcal D}^2 = {\mathcal D}_Q(2 - {\mathcal D}_Q)$, and thus has
its origin in UQSD.

A subtle point needs to be mentioned here. For $N > 2$, UQSD works only
when the states $\{|d_i\rangle\}$ are linearly independent. However,
that does not mean that the distinguishability defined by ${\mathcal D}_Q$ or
that by (\ref{Dtqas}) breaks down. The quantity ${\mathcal D}_Q$ 
can still be used as an upper bound on the probability with which the
linearly {\em dependent} states $\{|d_i\rangle\}$ can be unambiguously
distinguished. However, this discrimination may now be possible only in limited
situations, and the upper bound of the probability may not be achievable.
For example, for an orthonormal set of states $|a_1\rangle, |a_2\rangle,\dots,
|a_{n-1}\rangle$, if $|d_i\rangle = |a_i\rangle$, for $i=1, 2,\dots,n-1$,
and $|d_n\rangle = (|a_{n-2}\rangle+|a_{n-1}\rangle)/\sqrt{2}$, the
path-detector states $\{|d_i\rangle\}$ form a linearly dependent set, and
UQSD is not possible. But it is obvious that the states $|d_1\rangle$
through $|d_{n-3}\rangle$ can be unambiguosly distinguished, but not
$|d_{n-2}\rangle,|d_{n-1}\rangle,|d_{n}\rangle$.

\section{Wave-particle duality}
\subsection{Coherence and the duality relation}

Next we come to the wave nature of the quanton. For that we adopt the
coherence measure ${\mathcal C}$ used by Bera et al.  which is just the
normalized quantity ${\mathcal C}(\rho)={1\over N-1}\sum_{i\ne j} |\rho_{ij}|$,
where $\rho$ is the density operator of the quanton, and $\rho_{ij}$ are its
matrix elements in a particular basis. We assume that
${\mathcal C}$, calculated in the basis states $\{|\psi_i\rangle\}$ adequately
captures the wave-nature of the quanton. Recently it has been shown that
${\mathcal C}$ can actually be measured in an interference experiment
\cite{tania}. In this sense, it can be accorded the same status as the
fringe visibility.  In the presence of a path-detector, one
first has to trace over the states of the path-detector, to get a reduced
density matrix for the quanton. Doing that for the state (\ref{corrstate}),
one gets 
\begin{equation}
\rho_r = Tr_d[|\Psi\rangle\langle\Psi|]
 = \sum_{i=1}^n\sum_{j=1}^nc_ic_j^* \langle d_j|d_i\rangle |\psi_i\rangle\langle\psi_j|.
\end{equation}
Coherence can now be calculated using this reduced density matrix 
\begin{equation}
{\mathcal C} = {1\over N-1}\sum_{i\neq j} |c_ic_j| |\langle d_i|d_j\rangle|.
\label{C}
\end{equation}
This is the coherence of the quanton which has passed through the N-slit,
and is correlated with the path-detector with states $\{|d_i\rangle\}$.

Now, from (\ref{Dtqas}) and (\ref{C}) it is straightforward to see that
\begin{equation}
 {\mathcal D}^2 + {\mathcal C}^2 = 1 .
\label{ndualitye}
\end{equation}
This is a new duality relation for $N-$path interference, which restricts
the path distinguishability
and coherence. Notably, when the quanton and path-detector are in the pure
state (\ref{corrstate}), it is an
equality, and not an inequality. Hence the relation is tight. If other
experimental factors and mixedness are taken
into account, the coherence will only be smaller, and the duality relation
will in general be an inequality.

For $N=2, |c_1| = |c_2| = 1/\sqrt{2}$, we find
${\mathcal D} = \sqrt{1 - |\langle d_1|d_2\rangle|^2}$ and
${\mathcal C} = |\langle d_1|d_2\rangle|$.
For this case, ${\mathcal C}$ is exactly equal to Englert's visibility
${\mathcal V}$. Hence, ${\mathcal D}^2 + {\mathcal C}^2 = 1$ reduces to
${\mathcal D}^2 + {\mathcal V}^2 = 1$. It is also worth noting that
since ${\mathcal D}^2 = {\mathcal D}_Q(2 - {\mathcal D}_Q)$, the new duality
relation (\ref{ndualitye}) is mathematically equivalent to ${\mathcal D}_Q + {\mathcal C} = 1$
derived by Bera et al. \cite{nslit}.

The duality relation of Bera et al. has also been shown to hold for the
case where the initial state of the quanton and path-detector is not pure,
but mixed. However, in that case it is not an equality, but an inequality,
${\mathcal D}_Q + {\mathcal C}\le 1$. Consequently our {\em new duality relation
also holds for mixed states}, and becomes the inequality
${\mathcal D}^2 + {\mathcal C}^2 < 1$.
Thus, (\ref{nduality}) represents the most general duality relation for
$N-$path interference, which holds for pure and mixed states, and has
the form of Englert's duality relation (\ref{englert}). It is
saturated for pure states of quanton and path-detector.

\subsection{Path-distinguishability of Bagan et al.}

We now turn our attention to a recent work of Bagan et al. where a new
definition of distinguishability for multi-path was proposed, based on minimum error 
discrimination of states \cite{bagan}. To avoid confusion, we denote their
distinguishability by ${\mathcal D}_B$. The distinguishability of Bagan et al.
can be written as \cite{bagan}
\begin{equation}
{\mathcal D}_B \leq {1\over N-1}\sum_{i,j=1}^N \sqrt{\left({p_i+p_j\over 2}\right)^2
 - p_ip_j|\langle d_i|d_j\rangle|^2},
\label{DB}
\end{equation}
where $p_i$ should be identified with $|c_i|^2$ of our notation. The coherence
${\mathcal C}$ they have used is exactly the same as that in (\ref{C}). Their
duality relation can thus be written as ${\mathcal D}_B^2 + {\mathcal C}^2 \le 1$.
Since this relation has the same form of (\ref{nduality}), and
${\mathcal C}$ is the same in both, the {\em upper bound of} ${\mathcal D}_B$
should be directly comparable to the ${\mathcal D}$ given by (\ref{Dtqas}). 

For the special case $N=2$, (\ref{DB}) is an equality, it is easy to verify that
 ${\mathcal D}_B = {\mathcal D} = \sqrt{1 - 4p_1p_2|\langle d_1|d_2\rangle|^2}$.
For $N > 2$, it can be shown that the upper bound of ${\mathcal D}_B$ is not
the same as ${\mathcal D}$.
Since for the pure state (\ref{corrstate}), our distinguishability ${\mathcal D}$
saturates the inequality (\ref{nduality}), the upper bound of ${\mathcal D}_B$,
being different from
${\mathcal D}$, can only
be smaller than ${\mathcal D}$, and {\em will not saturate the inequality}
${\mathcal D}_B^2 + {\mathcal C}^2 \le 1$ of Bagan et al.  This shows that,
in general, the duality relation of Bagan et al. cannot be saturated for
$N > 2$, for pure entangled states of the quanton and path-detector.
Apart from this, the distinguishability of Bagan et al. has severe 
shortcomings, which we illustrate with some specific examples.

\begin{figure}
\centering
\resizebox{8.0cm}{!}{\includegraphics{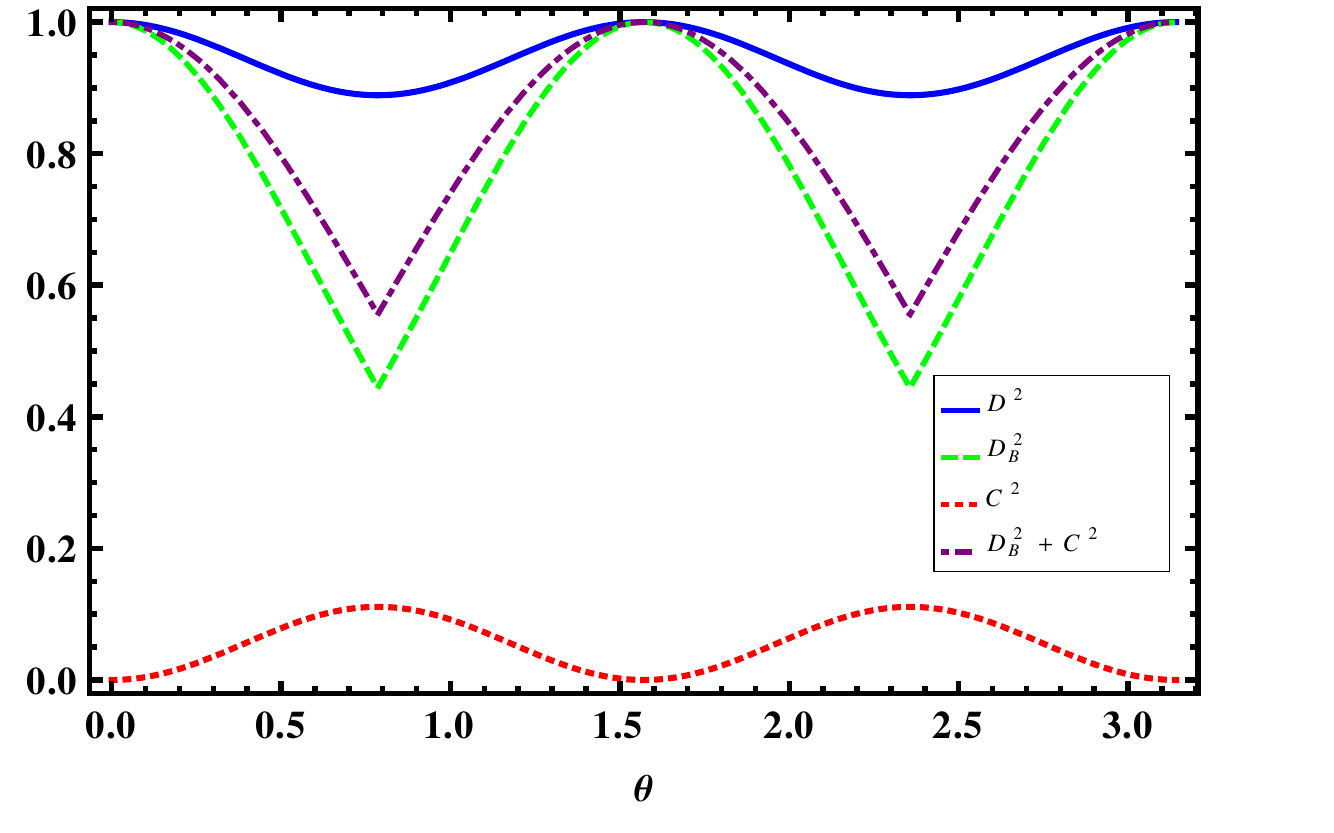}}
\caption{For $|d_1\rangle=\cos\theta|+\rangle+ \sin\theta|-\rangle$,
$|d_2\rangle=\sin\theta|+\rangle+ \cos\theta|-\rangle$,
and $|d_3\rangle=|0\rangle$, ${\mathcal D}^2$ (solid line),
the upper bound of ${\mathcal D}_B^2$ (dashed line), and ${\mathcal C}^2$ (dotted line) are plotted
against $\theta$. Clearly ${\mathcal D}$ is a stronger distinguishability
compared to the upper bound of ${\mathcal D}_B$, for the whole range of
$\theta$. In addition, the upper bound of
${\mathcal D}_B^2+{\mathcal C}^2$ (dot-dashed line) is not 1.
}
\label{DDB}
\end{figure}

Let us look at a specific case of $N=3$ where
$|d_1\rangle=\cos\theta|+\rangle+ \sin\theta|-\rangle$,
$|d_2\rangle=\sin\theta|+\rangle+ \cos\theta|-\rangle$,
and $|d_3\rangle=|0\rangle$, where $|+\rangle,|-\rangle,|0\rangle$ are 
orthonormal states, and $p_1=p_2=p_3=1/3$.
For this case, ${\mathcal D}^2= 1 - {1\over 9}\sin^22\theta$,
${\mathcal C}^2= {1\over 9}\sin^22\theta$, and
${\mathcal D}_B^2\leq {1\over 9}(2+\sqrt{1-\sin^22\theta})^2$.
From Figure \ref{DDB} one can see that ${\mathcal D}^2$ is very much larger than
the upper bound of  
${\mathcal D}_B^2$ for the whole range of $\theta$, except at points where it
saturates to 1. Our distinguishability
is clearly stronger than the distinguishability of Bagan et al. For this
case, the inequality of Bagan et al. is far from being saturated, as can
be seen from the plot of 
the upper bound of ${\mathcal D}_B^2+{\mathcal C}^2$. On the other hand,
${\mathcal D}^2+{\mathcal C}^2=1$, always (for pure states).

\begin{figure}
\centering
\resizebox{8.0cm}{!}{\includegraphics{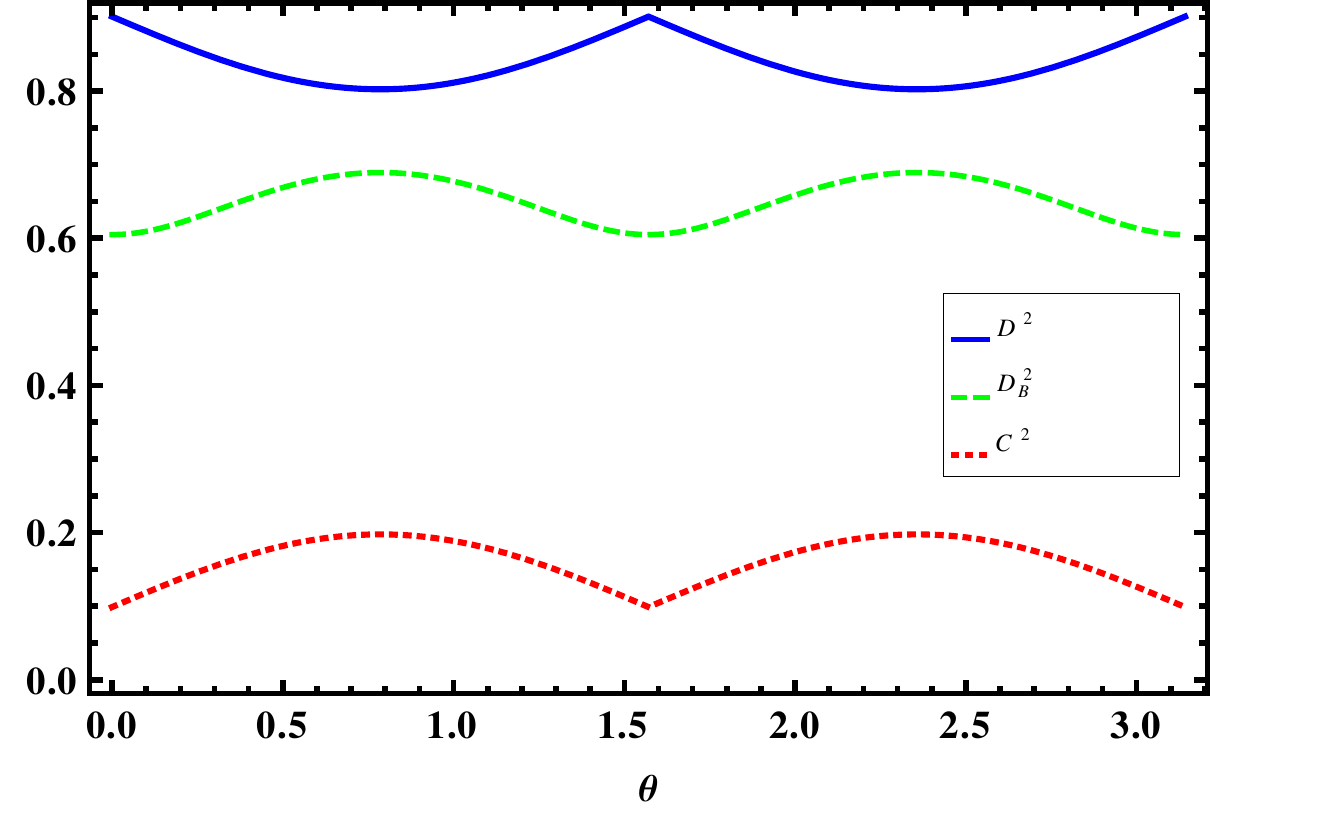}}
\caption{For $|d_1\rangle=\cos\theta|+\rangle+ \sin\theta|-\rangle$,
$|d_2\rangle=\sin\theta|+\rangle - \cos\theta|-\rangle$,
and $|d_3\rangle={2\sqrt{2}\over 3}|-\rangle+{1\over 3}|0\rangle$, and $p_1=p_2=p_3=1/3$, ${\mathcal D}^2$ (solid line), the upper bound of ${\mathcal D}_B^2$
(dashed line) and ${\mathcal C}^2$ (dotted line)
are plotted against $\theta$. Increasing ${\mathcal D}_B^2$ leads to increasing 
${\mathcal C}^2$ for the whole range of $\theta$. On the other hand, increasing ${\mathcal D}^2$ always leads to
decreasing ${\mathcal C}^2$. 
}
\label{DBC}
\end{figure}

Let us look at another specific case of $N=3$ where
$|d_1\rangle=\cos\theta|+\rangle + \sin\theta|-\rangle$,
$|d_2\rangle=\sin\theta|+\rangle - \cos\theta|-\rangle$,
$|d_3\rangle={2\sqrt{2}\over 3}|-\rangle+{1\over 3}|0\rangle$, and $p_1=p_2=p_3=1/3$.
Here $|d_1\rangle$ and $|d_2\rangle$ are orthogonal for all values of $\theta$.
For this case, ${\mathcal D}^2= 1 - {8\over 81}(|\sin\theta|+|\cos\theta|)^2$,
${\mathcal C}^2= {8\over 81}(|\sin\theta|+|\cos\theta|)^2$,
and\\
${\mathcal D}_B^2\leq {1\over 9}\left(1 + \sqrt{1-(8/9)\sin^2\theta}+\sqrt{1-(8/9)\cos^2\theta}\right)^2$. These
three quantities are plotted against $\theta$ (see Figure \ref{DBC}).
One can see that the upper bound of ${\mathcal D}_B^2$ and ${\mathcal C}^2$ increase and decrease together.
This clearly goes against the spirit of complementarity, as has also been
argued earlier in Ref. \cite{bimonte}.  When the initial entangled state of
the quanton and the path-detector is pure, any increase in path knowledge,
should lead to a decrease in the wave-aspect.
From (\ref{C}) one
can see that when $\mathcal{C}$ increases, the overlap of path-detector
states has to increase, which means the states are becoming less distinguishable.
Hence the distinguishability should go down. However Bagan et al.'s 
distinguishability increases, which
implies that the upper bound of ${\mathcal D}_B$ is not a good measure of path distinguishability.
Our path distinguishability ${\mathcal D}$, on the other hand, always satisfies
this criterion. 
Bagan et al. had mistakenly 
assumed that the duality relation of Bera et al., ${\mathcal D}_Q+{\mathcal C}=1$,
corresponds to the region below the line ${\mathcal D}_B+{\mathcal C}=1$, and thus misses a region
where wave and particle properties are compatible \cite{bagan}. The 
fact that for $N=2$, Bera et al.'s duality
relation reduces to Englert's relation (\ref{englert}) \cite{nslit},
shows that the assumption is not justified.

\section{Discussion}

To summarize, we have defined a new path-distinguishability ${\mathcal D}$ for 
$N-$path interference experiments, which is a measure of the particle nature
of a quanton. This distinguishability has its roots
in UQSD, and reduces to the distinguishability of Englert \cite{englert}
in the appropriate limit. A normalized coherence ${\mathcal C}$ has been 
proposed as a measure of the wave-nature of a quanton. It has recently been 
demonstrated that this coherence can be measured in an interference
experiment \cite{tania}. So it can be put on the same footing as interference
visibility.  For the 
two-slit case, this coherence reduces to the ideal interference
visibility, ideal in the sense that finite slit width etc. do not play
a role. The new path-distinguishability and coherence are shown to follow
a duality relation ${\mathcal D}^2+{\mathcal C}^2 \le 1$, which is saturated when
the entangled state of the quanton and path-detector is pure. This
duality relation can be treated as a quantitative statement of Bohr's
principle of complementarity, in the context of multi-path interference
experiments. It reduces to Englert's well-known duality relation for
two-slit interference in the appropriate limit, and has a same form.
Although the path-distinguishability ${\mathcal D}$ is based on UQSD, it
does not mean that the new duality relation (\ref{nduality}) holds only
for measurements of the UQSD kind. It represents a bound which should be
respected by any kind of error-free path-detection measurement. For pure
quanton-path-detector states, the duality relation is saturated to
the equality (\ref{ndualitye}). This indicates that for a duality relation of
the form (\ref{ndualitye}), for a given coherence, ${\mathcal D}$ represents
the strongest measure of path-distinguishability.

The distinguishability proposed by Bagan et al. \cite{bagan} does not satisfy
the expected criterion that the measures of particle nature and wave nature cannot
increase or decrease together. More precisely, we have shown that there are
cases when the overlap between the states of the path-detector decreases,
implying that the states should be more distinguishable, but the
distinguishability of Bagan et al. decreases, instead of increasing.
Failure of Bagan et al.'s approach to properly characterize
path-distinguishability indicates that minimum error discrimination
of states is probably not the right way to address the issue. UQSD appears to
provide a better answer.

\section*{Acknowledgements}
The authors thank Manabendra Bera and Arun Pati for useful discussions.
M.A. Siddiqui acknowledges financial support from the University Grants
Commission, India and Ramanujan Fellowship research grant (SB/S2/RJN-083/2014).

\end{document}